\documentclass[11pt]{article}
\usepackage{amssymb,latexsym}
\usepackage[dvips]{graphicx} 
\usepackage{url}

 \catcode`@=11\def\@oddhead{\hbox{}\hfil\rm\thepage}\def\@oddfoot{}
 \def\@evenhead{\hbox{}\hfil\rm\thepage}\def\@evenfoot{}
 \@twosidetrue \catcode`@=12

\DeclareFixedFont{\itshape}{OT1}{cmr}{m}{it}{11}

\newtheorem{prp}{Proposition}
\newtheorem{lem}[prp]{Lemma}\newtheorem{thm}[prp]{Theorem}
\newtheorem{cor}[prp]{Corollary}\newtheorem{cnj}[prp]{Conjecture}
\newenvironment{prf}{\begin{trivlist}\item[\emph{Proof.}]}{\end{trivlist}
  \medskip\par}
\newenvironment{prfof}[1]{\begin{trivlist}\item[\emph{Proof of #1.}]}{
  \end{trivlist} \medskip \par}
\newenvironment{rem}{\begin{trivlist}\item[\emph{Remarks.}]}{\end{trivlist}
  \medskip\par}
\def\prpb{\begin{prp}}\def\prpe{\end{prp}}
\def\lemb{\begin{lem}}\def\leme{\end{lem}}
\def\thmb{\begin{thm}}\def\thme{\end{thm}}
\def\corb{\begin{cor}}\def\core{\end{cor}}
\def\cnjb{\begin{cnj}}\def\cnje{\end{cnj}}
\def\prfb{\begin{prf}}\def\prfe{\end{prf}}
\def\prfofb#1{\begin{prfof}{#1}}\def\prfofe{\end{prfof}}
\def\remb{\begin{rem}}\def\reme{\end{rem}}
\def\prpa#1{\label{p:#1}}\def\prpu#1{Proposition~\ref{p:#1}}

\def\thma#1{\label{t:#1}}\def\thmu#1{Theorem~\ref{t:#1}}

\def\seca#1{\label{s:#1}}\def\secu#1{Section~\ref{s:#1}}

\def\itmb{\begin{enumerate}}\def\itme{\end{enumerate}}
\def\itdb{\begin{itemize}}\def\itde{\end{itemize}}
\def\ittb{\begin{description}}\def\itte{\end{description}}
\def\eqnb{\begin{equation}}\def\eqne{\end{equation}}
\def\arrb#1{\begin{array}{#1}}\def\arre{\end{array}}
\def\tabb#1{\par\noindent\begin{tabular}{#1}}
\def\tabe{\end{tabular}\par\noindent}
\def\eqna#1{\label{e:#1}}\def\eqnu#1{(\ref{e:#1})}

\def\QED{\relax\ifmmode\let\@tempa\relax\ifcase\@eqcnt\def\@tempa{& & &}\or
  \def\@tempa{& &}\else\def\@tempa{&}\fi\@tempa $\Box$ \else\hfill $\Box$ \fi}
\def\DDD{\relax\ifmmode\let\@tempa\relax\ifcase\@eqcnt\def\@tempa{& & &}\or
 \def\@tempa{& &}\else\def\@tempa{&}\fi\@tempa $\Diamond$
 \else\hfill $\Diamond$ \fi}

\def\Rom#1{\uppercase\expandafter{\romannumeral#1}}
\def\dsp{\displaystyle}

\def\AHFA{\hbox{\vrule height16pt depth10pt width0pt}\displaystyle}

\def\vec#1{{\mathbf{#1}}} \def\vec#1{\overrightarrow{#1}}
\def\clvec#1#2#3{\def\clvecone{#3}\left(\arrb{c} \dsp #1\\ \dsp #2
 \ifx\clvecone\empty\else\\ \dsp #3\fi\arre\right)}

\def\diff#1#2{\dsp\frac{d\,#1}{d#2}}

\def\pderiv#1#2{\dsp\frac{\partial\,#1}{\partial#2}}

\def\le{\leqq} \def\ge{\geqq}

\def\preals{{\mathbb R_+}}

\def\prb#1{\def\prbone{#1}
  \ifx\prbone\empty{\mathrm{P}}\else{\mathrm{P[\;}}#1{\mathrm{\;]}}\fi}
\def\prbseq#1#2{\def\prbseqone{#2}
  \ifx\prbseqone\empty{\mathrm{P}}_{#1}\ignorespaces
  \else{\mathrm{P}}_{#1}{\mathrm{[\;}}#2{\mathrm{\;]}}\fi}

\def\EEseq#1#2{\def\EEseqone{#2}
  \ifx\EEseqone\empty{\mathrm{E}}_{#1}\else
 {\mathrm{E}}_{#1}{\dsp\mathrm{[\;}}#2{\mathrm{\;]}}\fi}

\def\VVseq#1#2{\def\VVseqone{#2}
  \ifx\VVseqone\empty{\matrm{V}}_{#1}\else
 {\mathrm{V}}_{#1}{\dsp\mathrm{[\;}}#2{\mathrm{\;]}}\fi}

\def\figa#1{\label{f:#1}}\def\figu#1{Fig.~\ref{f:#1}}

\title{
 Equation of motion for incompressible mixed fluid driven by evaporation
and its application to online rankings
}
\author{
Kumiko Hattori
 \\ {\small
Department of Mathematics and Information Sciences,
 Tokyo Metropolitan University, 
} \\ {\small
Hachioji, Tokyo 192-0397, Japan.
} \\ {\small
email: \url{khattori@tmu.ac.jp}
}
\\ \and
Tetsuya Hattori
 \\ {\small
Mathematical Institute, Graduate School of Science, Tohoku University,
 Sendai 980-8578, Japan.
} \\ {\small
URL: \url{http://www.math.tohoku.ac.jp/hattori/amazone.htm}
} \\ {\small
email: \url{hattori@math.tohoku.ac.jp}
}
}
\date{\today}

\begin{document}
\maketitle

\begin{center}
ABSTRACT
\end{center}

We give a unique classical solution
to initial value problem for a system of partial differential equations
for the densities
of components of one dimensional incompressible fluid mixture
driven by evaporation.

Motivated by the known fact that the solution appears as 
an infinite particle limit of stochastic ranking processes,
which is a simple stochastic model of time evolutions of
e.g., Amazon Sales Ranks,
we collected data from the web 
and performed statistical fits to our formula.
The results suggest that the fluid equations and solutions 
may have an application in the analysis of online rankings.

\vspace*{0.5in}\par\noindent\textit{Keywords:} 
 evaporation driven fluid;
 non-linear wave;
 stochastic ranking process;
 long tail; Pareto distribution;

\vspace*{0.1in}\par\noindent\textit{2000 Mathematics Subject Classification:}
Primary 35C05; Secondary 35Q35, 82C22

\vspace*{0.1in}\par\noindent\textit{running head:}
Mixed fluid driven by evaporation and online rankings

\vspace*{0.2in}\par\noindent\textit{Corresponding author:} 
Tetsuya Hattori, 
\url{hattori@math.tohoku.ac.jp}
\par\noindent
Mathematical Institute, Graduate School of Science, Tohoku University,
 Sendai 980-8578, Japan
\par\noindent
tel+FAX:  011-81-22-795-6391

\newpage

\section{Introduction.}
\seca{1}

Let $f_i>0$, $i=1,2,\cdots$, be positive constants, and
consider the following system of non-linear partial differential equations
for the functions $u_i(y,t)$, $i=1,2,\cdots$, and $v(y,t)$,
defined on $(y,t)\in [0,1)\times \preals$:
\eqnb
\eqna{e1}
\pderiv{u_{i}(y,t)}{t} +\pderiv{(v(y,t)\, u_{i}(y,t))}{y}
=-f_{i} u_{i}(y,t),
\ \  i=1,2,\cdots,
\eqne
\eqnb
\eqna{e3}
\sum_{j} u_{j}(y,t)=1\,.
\eqne
We consider initial value problems for smooth non-negative initial data
\[
u_{i}(y,0)\ge0,\ \ 0\le y<1,\  i=1,2,\cdots,
\]
with the boundary conditions at $y=0$ and $y=1$:
\eqnb
\eqna{bc1}
v(1-0,t)=0,
\eqne
\eqnb
\eqna{bc2}
u_{i}(0,t)= 
\frac{\dsp f_{i}\rho_{i}}{
\dsp \sum_{j}f_{j}\rho_{j}}\,,
\ \  i=1,2,\cdots,
\eqne
for $t\ge 0$, where, for each $i$,
\eqnb
\eqna{rho}
 \rho_{i}=\int_0^1 u_{i}(z,0)\, dz,
\eqne
and we assume
\eqnb
\eqna{ellone}
\sum_j f_j\rho_j <\infty.
\eqne
Note that adding up \eqnu{e1} over $i$ and applying \eqnu{e3} we have
\eqnb
\eqna{e2}
\pderiv{v(y,t)}{y}
=-\sum_{j} f_{j} u_{j}(y,t),
\eqne
which, with \eqnu{bc1}, determines $v$ in terms of $u_i$\,.

Given the positive constants $f_i$ and the initial data $u_i(y,0)$,
the set of equations \eqnu{e1} \eqnu{e3} \eqnu{bc1} \eqnu{bc2}
defines the evolution of our system.
The following arguments and results hold 
both for finite components ($i=1,2,\cdots,N$) and infinite components.
(In fact, we can extend the system and the solution to a case with
any probability space $\Omega$, by replacing $u_i(y,t)$ with 
a measure $\mu(d\omega,y,t)$.
See \cite{HH071} for probability theoretic arguments.)

A physical meaning of the system is as follows.
We are considering a motion of incompressible fluid mixture
in an interval with length normalized to $1$, where
$u_i(y,t)$ is the density of $i$-th component at space-time point 
$(y,t)$.
\eqnu{e3} implies that we normalize $u_i$ so that it represents the
ratio of $i$-th component.
We are naturally interested in the non-negative solutions $u_i(y,t)\ge 0$.
$v(y,t)$ is the velocity field of the fluid.
\eqnu{e1} is the equation of continuity, and the right-hand-side implies
that each component evaporates with rate $f_i$ per unit time and unit mass.
Note that
the set of equations \eqnu{e2} and \eqnu{bc1} is equivalent to
\eqnb
\eqna{velocity}
v(y,t) =\sum_j f_j\, \int_{y}^1 u_j(z,t) dz,
\eqne
which implies that the velocity field, or the motion of the fluid,
is caused solely by filling the amount of fluid which evaporated
from the right side of the point $y$.
In particular, we have no flux through the boundary $y=1$ ($v(1-0,t)=0$).

The boundary condition \eqnu{bc2} at $y=0$ is so tuned by the initial data
that the loss of mass by evaporation is compensated by the immediate
re-entrance at $y=0$ as liquidized particles, 
so that the total mass of each fluid component in the interval $[0,1)$
is conserved over time:
\eqnb
\eqna{e4}
\int_0^1 u_{i}(z,t)\,dz= \rho_{i},\ \ t\ge0.
\eqne
In fact, 
\eqnu{e4} is equivalent to \eqnu{bc2},
under the conditions \eqnu{e1} \eqnu{e3} \eqnu{bc1} \eqnu{rho},
if $\sup_i f_i<\infty$.
(See Appendix~\ref{s:appendix}.)

Let
\eqnb
\eqna{ytildeCyt}
y_C(y,t)=1-\sum_j e^{-f_j t} \int_y^1 u_j(z,0)\,dz,
\ \ 0\le y< 1,\ t\ge0.
\eqne
For each $t\ge 0$,
$y_C(\cdot,t):\ [0,1)\to[y_C(0,t),1)$ is a continuous, 
strictly increasing,
onto function of $y$, and  
its inverse function
$\dsp \hat{y}(\cdot,t):\ [y_C(0,t),1)\to[0,1)$ exists:
\eqnb
\eqna{yhatyt}
1-y=\sum_j e^{-f_j t} \int_{\hat{y}(y,t)}^1 u_j(z,0) \,dz,
\ \ y_C(0,t)\le y< 1,\ t\ge0.
\eqne
$y_C(y,t)$ denotes the position of a fluid particle at time $t$
(on condition that it does not evaporate up to time $t$)
whose initial position is $y$.
$\hat{y}(y,t)$ denotes the initial position of a fluid particle
located at $y$ ($\ge y_C(0,t)$) at time $t$.

With slight abuse of notations,
we will often write $y_C(t)$ for $y_C(0,t)$:
\eqnb
\eqna{yCtdiscrete}
y_C(t)=y_C(0,t)=1-\sum_{j}\rho_{j}e^{-f_{j}  t}.
\eqne
$\dsp y_C:\ [0,\infty)\to[0,1)$ is a continuous, strictly increasing,
onto function of $t$, and its inverse function
$\dsp t_0:\ [0,1)\to [0,\infty)$ exists:
\eqnb
\eqna{t0y}
y_C(t_0(y))=y, \ \ 0\le y<1\,.
\eqne
In this paper, we prove the following.
\thmb
\thma{solution}
There exists a unique classical solution 
to the initial value problem for the system
of partial differential equations defined by \eqnu{e1}--\eqnu{rho},
which is explicitly given by \eqnu{velocity} and
\eqnb
\eqna{Tets20070726a}
u_{i}(y,t)  =\left\{\arrb{ll}\dsp
\frac{\dsp
e^{-f_{i} t_0(y)} f_{i} \rho_{i} 
}{\dsp
\sum_{j} e^{-f_{j} t_0(y)} f_{j} \rho_{j} 
}\,, & y<y_C(t), \\
\frac{\dsp
e^{- f_{i} t} u_{i}(\hat{y}(y,t),0 ) 
}{\dsp
\sum_{j} e^{-f_{j} t} u_{j}(\hat{y}(y,t),0) 
}\,, & y>y_C(t),
\arre \right. 
\hspace*{3cm} \mbox{\DDD}
\eqne
\thme
Note the unique feature of the solution that
for $y<y_C(t)$ the solution is stationary:
\eqnb
\eqna{dudt0}
\pderiv{u_i}{t}(y,t)=0,\ \ y<y_C(t),
\eqne
while initial conditions affect $y>y_C(t)$ part only
through wave propagation.

In natural phenomena where evaporation is active,
such as producing salt out of sea water,
viscosity, surface tension, and
external forces such as gravitational forces dominate,
and the effect of evaporation on the motion of fluid
would be relatively too small to observe.
Thus the equation and the solution we consider in this paper 
may not have attracted much attention.
However, 
there are phenomena on the web for which our formulation
may work as a simplified mathematical model,
such as the time evolutions of rankings of book sales in 
the online booksellers.
Such possibility is theoretically based on a result
that \eqnu{Tets20070726a} appears as an infinite particle limit
of the stochastic ranking process  \cite{HH071}, which is
a simple model of the time evolutions of e.g., 
the number known as the Amazon Sales Rank.
(We note that this number has mathematically little to
do with the perhaps more popular notion of Google Page Ranks.)
We collected data of the time evolution of the numbers from the web,
and performed statistical fits of the data to \eqnu{yCtdiscrete}.
Considering the simplicity of our model and formula, the fits seem good,
which suggest that there is a new application of our results
in the analysis of online rankings.

The plan of this paper is as follows.
In \secu{solution} we give a proof of \thmu{solution}.
In \secu{webranking} we give results of 
fits to \eqnu{yCtdiscrete} of data from the web.

\smallskip\par 
The authors would like to thank Prof.~T.~Miyakawa
and Prof.~M.~Okada for
taking interest in our work and for inviting the authors to their seminars.
The research of K.~Hattori 
is supported in part by a Grant-in-Aid for 
Scientific Research (C) 16540101 from the Ministry of Education,
 Culture, Sports, Science and Technology, and
the research of T.~Hattori 
is supported in part by a Grant-in-Aid for 
Scientific Research (B) 17340022 from the Ministry of Education,
 Culture, Sports, Science and Technology.

\section{Proof of the main theorem.}
\seca{classicstationaryexp}
\seca{solution}

Put
\eqnb
\eqna{spacialdistributionfcn}
U_i(y,t)=\int_y^1 u_i(z,t)\,dz,\ \ i=1,2,\cdots.
\eqne
With \eqnu{spacialdistributionfcn},
the the system of equations \eqnu{e1}--\eqnu{rho} is equivalent to
the following:
$U_i(y,t)$ is decreasing in $y$ and $U_i(1-0,t)=0$, and
\eqnb
\eqna{e1i}
\pderiv{U_{i}}{t}(y,t) +v(y,t)\, \pderiv{U_{i}}{y}(y,t)
=-f_{i} U_{i}(y,t),
\ \  i=1,2,\cdots,
\eqne
\eqnb
\eqna{e3i}
\sum_{j} U_{j}(y,t)=1-y\,,
\eqne
\eqnb
\eqna{velocityi}
v(y,t) =\sum_j f_j\, U_j(y,t),
\eqne
for $0\le y<1$, $t\ge0$ (note \eqnu{velocity}), and 
noting \eqnu{e4},
\eqnb
\eqna{bc2i}
U_{i}(0,t)= \rho_i
\ \  i=1,2,\cdots,\ t\ge 0.
\eqne

Now, for each $0\le y_0<1$ let $y_B=y_B(y_0;t)$ be a 
solution to an ODE
\eqnb
\eqna{yB}
\diff{y_B}{t}(y_0;t)=v(y_B(y_0;t),t),\ t\ge0,
\ \ y_B(y_0;0)=y_0\,,
\eqne
and put
\eqnb
\eqna{phi}
\phi_i(t)=U_i(y_B(y_0;t),t).
\eqne
With \eqnu{yB} and \eqnu{e1i} it follows that
\[
\diff{\phi_i}{t}(t)=\pderiv{U_i}{t}(y_B(y_0;t),t)
+v(y_B(y_0;t),t)\,\pderiv{U_i}{y}(y_B(y_0;t),t)
=-f_{i} U_{i}(y_B(y_0;t),t)=-f_i \phi_i(t),
\]
hence, with $\phi_i(0)=U_i(y_0,0)$,
$\phi_i$ is uniquely solved as
\eqnb
\eqna{phisol}
\phi_i(t)=U_i(y_0,0)\, e^{-f_it}.
\eqne
With \eqnu{velocityi} and \eqnu{yB}, we then find
\[
\diff{y_B}{t}(y_0;t) = \sum_j f_j U_j(y_0,0)\, e^{-f_j t},
\]
hence, using \eqnu{e3i} and \eqnu{spacialdistributionfcn}, we have
\eqnb
\eqna{yBsol}
y_B(y_0;t) = 1- \sum_j U_j(y_0,0)\, e^{-f_j t}
=1- \sum_j \int_{y_0}^1 u_j(z,0)\,dz\, e^{-f_j t}
=y_C(y_0,t),
\eqne
where $y_C$ is defined in \eqnu{ytildeCyt}.
With \eqnu{yhatyt}, \eqnu{phi} and \eqnu{phisol} we uniquely obtain
\[
U_i(y,t)=U_i(\hat{y}(y,t),0)\, e^{-f_it}.
\]
Differentiating by $y$ and using 
\eqnu{spacialdistributionfcn} and \eqnu{yBsol}
\[
u_i(y,t)=\left(\pderiv{y_B}{y_0}(\hat{y}(y,t);0)\right)^{-1}
u_i(\hat{y}(y,t),0)\, e^{-f_it}
=\frac{u_i(\hat{y}(y,t),0)\, e^{-f_it}}{\dsp
\sum_j u_j(\hat{y}(y,t),0)\, e^{-f_j t}}\,,
\]
where $\dsp \pderiv{y_B}{y_0}$ is the derivative of $y_B=y_B(y_0;t)$
with respect to the parameter $y_0$.
This proves \eqnu{Tets20070726a} for $y>y_C(t)$.

Next let $y<y_C(t)$ and put $t_1=t-t_0(y)\in (0,t)$, where $t_0$ is
as in \eqnu{t0y}.
Let $y_A$ be a solution to an ODE
\eqnb
\eqna{yA}
y_A'(s)=v(y_A(s),s),\ s\ge t_1\,,
\ \ y_A(t_1)=0\,,
\eqne
and put
\eqnb
\eqna{phiA}
\phi_i(s)=U_i(y_A(s),s), \ s\ge t_1\,.
\eqne
Note that \eqnu{bc2i} implies $\phi_i(t_1)=U_i(0,t_1)=\rho_i$,
hence, as below \eqnu{yB}, 
$\phi_i$ is uniquely solved as
\eqnb
\eqna{phiAsol}
\phi_i(s)=\rho_i\, e^{-f_i(s-t_1)}.
\eqne
With \eqnu{velocityi} and \eqnu{yA}, we then find
\[
y_A'(s) = \sum_j f_j \rho_j\, e^{-f_j(s-t_1)},
\]
Note that \eqnu{rho} and \eqnu{e3} imply $\dsp \sum_j \rho_j=1$.
Hence,
\eqnb
\eqna{yAsol}
y_A(s) = 1- \sum_j \rho_j\, e^{-f_j(s-t_1)}
=y_C(s-t_1),\ s\ge t_1\,.
\eqne
where $y_C$ is defined in \eqnu{yCtdiscrete}.

Putting $s=t$ in \eqnu{phiA} and \eqnu{phiAsol}, 
using \eqnu{yAsol}, and recalling that $t_1=t-t_0(y)$,
we have, with \eqnu{t0y},
\[
U_i(y,t)=U_i(y_C(t_0(y)),t)=\rho_i\, e^{-f_i t_0(y)}.
\]
Differentiating by $y$ and using 
\eqnu{spacialdistributionfcn}, \eqnu{t0y} and \eqnu{yCtdiscrete},
\[
u_i(y,t)=
\left(\diff{y_C}{t}(t_0(y))\right)^{-1}\rho_if_i \, e^{-f_i t_0(y)}
=\frac{\rho_if_i \, e^{-f_i t_0(y)}}{\dsp
\sum_{j}\rho_{j}f_j e^{-f_{j}  t_0(y)}}\,,
\]
which proves \eqnu{Tets20070726a} for $y<y_C(t)$.
This completes a proof of \thmu{solution}.

\bigskip\par
Before closing this section,
we give a couple of examples of the solution to the equation of motion.

\paragraph*{Example 1 (One particle type).}
For a pure fluid, $\rho_1=1$, hence we have
$u_1(y,t) \equiv 1$.
Since there is only one type of incompressible fluid,
the density is constant.
However, there is a flow driven by evaporation
even in this case, and we actually have
$\dsp y_C(t)=1-e^{-f_1 t}$.

\paragraph*{Example 2 (Two particle types with $f_2 =0$).}
Consider a mixture of $2$ components with the ratios
satisfying $\rho_1>0$ and $\rho_2=1-\rho_1>0$.
$f_2=0$ means no evaporation, 
so the situation is a salty sea with salt density $\rho_2$,
where evaporation and flow from a river balance.
In \thmu{solution} we assumed $f_2>0$,
but we can calculate explicitly for $f_2\ge 0$.
We have
\[
y_C(t)=\rho_1(1-e^{-f_1t}),\ \ \ 
t_0(y)=-\frac{1}{f_1}\log (1-\frac{y}{\rho_1}).\]
The expressions of $u_i(y,t)$ for $y<y_C(t)$ become simple:
\[
v(y,t)=f_1\, (\rho_1-y),\ \ \ 
u_1 (y,t)  \equiv 1, \ \  u_2 (y,t)  \equiv 0 \,.
 \]
The pure water from the river comes in up to $y<y_C(t)$.
(Note that we are considering a fictitious $1$-dimensional case
where no spacial mixing such as turbulence occurs and we have no other
dynamics such as diffusion.)

If, furthermore, the initial distribution is uniform on $[0,1)$:
$u_{i}(y,0)=\rho_{i}$, $i=1,2$,  
then the expressions for $y>y_C(t)$ are also simple:
\[
y_C(y,t)=1-(1-y) (\rho_1 e^{-f_1 t}+ \rho_2 ),
\ \ \hat{y}(y,t)=
1-\frac{1-y}{\rho_1 e^{-f_1 t}+ \rho_2}\,,
\]
and consequently,
\[
 v(y,t)=(1-y)f_1 \frac{\rho_1 e^{-f_1 t}}{\rho_1 e^{-f_1 t}+ \rho_2},
\ \ \ u_1(y,t)=\frac{\rho_1 e^{- f_1 t} }{\rho_1 e^{-f_1 t}+ \rho_2}\,,
\ \ u_2(y,t)=\frac{\rho_2}{\rho_1 e^{-f_1 t}+ \rho_2}\,, 
\]
 for $y>y_C(t)$.
(In general, the formulas are dependent on initial data in complex ways,
and we do not have explicit formula.)

\section{Possible application to rankings on the webs.}
\seca{webranking}

\subsection{Pareto distribution.}
\seca{classicstationaryexpPareto}
\seca{Pareto}

An idea that the time evolution of ranking numbers
such as in Amazon.co.jp sales rank may be explained by 
evaporation-driven fluid motion, is theoretically
based on a result \cite{HH071} which says that
\eqnu{Tets20070726a} appears as a time evolution of
empirical distributions of the stochastic ranking processes 
in an infinite particle limit.
Roughly speaking, the stochastic ranking process
is a simple model of the time evolution of 
a list of rankings, such as 
a ranking of book sales in an online bookseller,
or a table of page titles in a collected web bulletin board.
A particle (a book, in the case of online booksellers)
in the ranking jumps randomly to the rank $1$ (each time the book is sold),
and increases the ranking number by $1$ each time some other particle 
of larger ranking number jumps to rank $1$.
Because an increase in ranking number 
is a result of jumps of very large number of particles
in the tail side of the ranking,
the particle effectively moves on the ranking queue in a deterministic way,
even though each jump occurs at a random time.
The jump corresponds to the evaporation in the fluid model.
We can therefore predict the time evolutions of 
rankings appearing on the web rankings based on our model.
See \cite{HH071} for details of the stochastic ranking processes.

Here we try to see how a trajectory of a particle \eqnu{yCtdiscrete}
could be observed in the web rankings.
In applying \eqnu{yCtdiscrete} to the actual rankings,
we need to choose the distribution of evaporation rates
$\{(f_i, \rho_i)\mid i=1,2,\cdots,N\}$.
In the case of social or economic studies such as online booksellers,
this corresponds to choosing the distribution of activities or transactions.
In the case of online booksellers, we have to chose the
distribution of sales rates over books.
The Pareto distribution (also called log-linear distribution in social
studies, or power-law in physics literatures) is traditionally used 
as a basic model distribution for various social rankings,
perhaps a most well-known example is the ranking of incomes.
Let $N$ be the total size of population, and for $i=1,2,\cdots,N$,
denote by $f_i$ the income of the $i$-th wealthiest person.
If
\eqnb
\eqna{Paretodiscrete}
f_i=a\left(\frac{N}{i}\right)^{1/b},\ \rho_i=\frac1N\,,\ \ i=1,2,3,\cdots,N,
\eqne
holds for some positive constants $a$ and $b$,
then the distribution of incomes is said to satisfy the Pareto distribution.
(The Pareto distribution assumes all the constituents to have
distinct $f_i$\,, which leads to equal weight $\rho_i=1/N$ in our notation.)
The constant $a$ corresponds to the smallest income, and
the exponent $b$ reflects a social equality of incomes:
in fact the ratio of the largest income to the smallest is $f_1/f_N=N^{1/b}$,
which is close to $1$ if $b$ is large (a fair society), while is large
(society is in monopoly) if $b$ is small.
(Our $b$ corresponds to $\alpha$ in a standard textbook on statistics,
$\theta$ in \cite{CG}, and $-1/\beta_2$ in \cite{BSH}.)

Substituting \eqnu{Paretodiscrete} in \eqnu{yCtdiscrete} of
\secu{classicstationaryexp},
and approximating the summation by integration, we have,
after a change of variable,
\eqnb
\eqna{yCPareto}
y_C(t)=1-b(at)^b \Gamma(-b,at) +O(N^{-1}),
\eqne
where
$\dsp \Gamma(z,p)=\int_{p}^{\infty} e^{-w} w^{z-1} \, dw$
is the incomplete Gamma function.
$y_C(t)$ is a relative ranking normalized by $N$,
so the time evolution of ranking $x_C(t)$ is
\eqnb
\eqna{y2x}
x_C(t)=1+N\, y_C(t).
\eqne
The $O(N^{-1})$ contribution in \eqnu{yCPareto} is 
(by a careful calculation) seen to be non-negative and bounded by
 $\dsp \frac1{N}e^{-at}\le \frac1{N}$, leading to a difference of
 at most $1$ in the ranking $x_C(t)$, which is insignificant
for our applications below, so we will ignore it.

Note that $\Gamma(-b,at)\to\infty$ as $t\to 0$, for $b>0$.
This divergence is harmless because it is cancelled by $t^b$ 
in \eqnu{yCPareto},
but for numerical and asymptotic analysis,
it is better to perform a partial integration on the right-hand side
to find
\eqnb
\eqna{classicstationaryexpPareto}
y_C(t)=1-e^{-at}+ (at)^b \Gamma(1-b,at),
\eqne
which, with \eqnu{y2x}, leads to
\eqnb
\eqna{xclassicstationaryexpPareto}
x_C(t)=N\,(1-e^{-at}+ (at)^b \Gamma(1-b,at)) + 1.
\eqne
The constant $a$, which denotes the lowest income
in the Pareto distribution, has a role of a time constant 
in \eqnu{xclassicstationaryexpPareto}.
In particular, the short time behavior of $x_C(t)$ for $0<b<1$ is
\eqnb
\eqna{classicalstationaryParetoshorttimex}
x_C(t)=  c\, t^b+O(t),
\eqne
where
\eqnb
\eqna{Na2c}
c= Na^b \Gamma(1-b).
\eqne
For $1<b<2$ we need a partial integration once more for a better expression;
\eqnb
\eqna{classicstationaryexpPareto15}
y_C(t)=1-e^{-at}(1-\frac{at}{b-1})
- \frac{(at)^b}{b-1}\, \Gamma(2-b,at)
=\frac{ab}{b-1}\,t - \Gamma(2-b)\,\frac{a^b}{b-1}\,t^b +O(t^2).
\eqne
Note that for $0<b<1$ the leading short time
behavior is $y_C(t)=O(t^b)$, which is tangential to the $y$ axis at $t=0$,
while for $b>1$ (the case $b\ge2$ can be handled similarly)
the linear dependence $y_C(t)=O(t)$ is dominant for small $t$.

\subsection{2ch.net bulletin board thread index listings.}
\seca{2ch}

2ch.net is one of the largest collected web bulletin boards in Japan.
Each category (`board') has an index listing of the titles of `threads' or 
the web pages in the board.
The titles are ordered by ``the last written thread at the top'' principle;
if one writes an article (`response') to a thread, the title of that
thread in the index listing jumps to the top instantaneously, and
the titles of other threads which were originally nearer to the top
are pushed down by $1$ in the listing accordingly.
We can extract the exact time that a thread jumped to rank $1$,
because the time of each response in a thread is recorded together
with the response itself.
All these features of the 2ch.net index listing match the definition
of the stochastic ranking process in \cite{HH071}.

As a first attempt to apply our theoretical results to online data,
we collected data of the time evolution of the index listing
and performed statistical fits of the data to \eqnu{yCtdiscrete}.
The actual properties of jump rates would be more involved than the
models; for example, the distribution of jump rates 
(namely, the distribution of the frequencies of responses of the threads)
would be more complex than the Pareto distribution,
and looking into actual data
would provide a test to the applicability of our simple model.

We note that we can use deterministic (non-stochastic)
formula such as \eqnu{xclassicstationaryexpPareto} if $N$,
the total number of threads in the board is large.
When we are keeping track of the ranking of a thread,
it goes down (number increases) when and only when other threads at
the tail side of the ranking jumps to the top (i.e., someone writes
a response
to one of these other threads), and though each jump occurs randomly,
since there are $O(N)$ threads on the tail side of the thread
in question (unless it is extremely near the tail),
a law-of-large-numbers like mechanism works 
(as rigorously proved in \cite{HH071}), and the time evolution
of the ranking for each thread becomes deterministic
as in \eqnu{xclassicstationaryexpPareto}.
In the case of 2ch.net, $N$ is roughly about $700$ to $800$, so
we would expect fluctuation of a few percent,
and up to that accuracy, we expect a time evolution predicted by
\eqnu{xclassicstationaryexpPareto}.

We also note that the time evolutions are independent of which
thread one is looking at, because the changes in the ranking
are caused by the collective motion of the threads towards the tail;
popular threads jump back to the top ranking more frequently than 
the less popular ones,
but as long as the threads remain in the queue (i.e., before the next jump),
both a popular thread and an unpopular thread should behave in the same way,
depending only on their position in the ranking.

\begin{figure}[hbt]
\begin{center}
\includegraphics[scale=0.58]{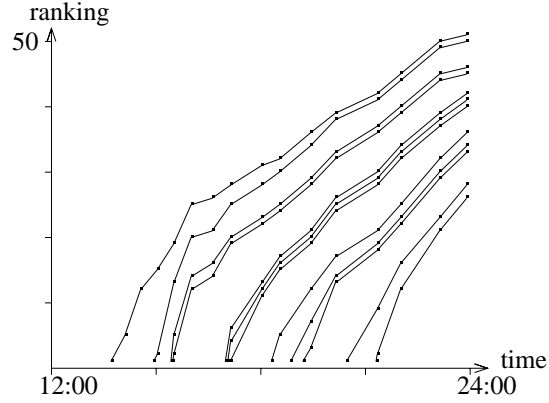}
\caption{Record of ranking changes in an afternoon for
 $12$ threads in a board of 2ch.net. Points from a thread are
joined by line segments to guide the eye.}
\figa{2ch_aa}
\end{center}
\end{figure}
\figu{2ch_aa} is a plot of the threads in a board
which jumped to the rank $1$
during active hours one day and stayed in the queue
without jumps until midnight. 
There were $12$ such threads.
The ranking is obviously monotone function of time between jumps,
and there are no overtaking, so that the lines do not cross in the figure.
\begin{figure}[hbt]
\begin{center}
\includegraphics[scale=0.55]{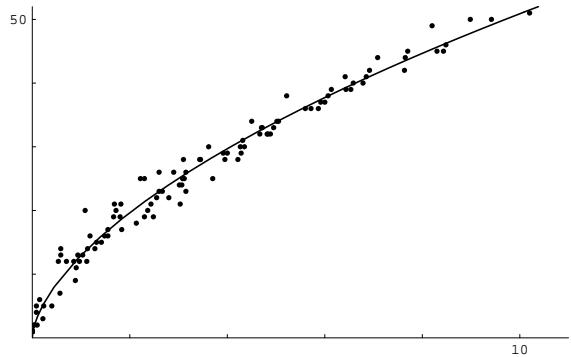}
\caption{Collection of $12$ threads in a board of 2ch.net, same as in
\protect\figu{2ch_aa}. For each thread,
time is shifted so that the rank of the thread is $1$ at time $0$.
The curve is 
$x_C(t)$ of \protect\eqnu{xclassicstationaryexpPareto}
with the best fit $a=a^*$ and $b=b^*$ to the data.
Horizontal and vertical axes are the hours and ranking, respectively.}
\figa{2ch_bb}
\end{center}
\end{figure}
\figu{2ch_bb} is a plot of same data as in \figu{2ch_aa},
except that, for the horizontal axis the time is so shifted
for each thread that the ranking of the thread is $1$ at time $0$.
Though each thread starts at rank $1$ on different time of the day,
\figu{2ch_bb} shows that time evolutions after rank $1$ are
on a common curve.
$N$ is the total number of threads, which is
 $N=795$ at the time of observation for \figu{2ch_bb},
and $a$ and $b$ are positive constant parameters
to be determined from the data.
We performed a least square fit 
to \eqnu{xclassicstationaryexpPareto}
of $n_d=117$ data points shown in \figu{2ch_bb}.
The best fit for the parameter set $(a,b)$ is
$(a^*,b^*)=(3.3425\times 10^{-4}, 0.6145)$
($\sqrt{\chi^2/n_d}\simeq 1.8$).
In particular, we see a rather clear behavior close to the origin
that the plotted points are on a curve tangential to $y$ axis, indicating 
 $x_C(t)=O(t^b)$ with $b<1$
 as in \eqnu{classicalstationaryParetoshorttimex}.
Considering the simplicity of our model and formula, 
the fits seem good,
suggesting a possibility of new application of fluid dynamics
in the analysis of online rankings.

\subsection{Amazon.co.jp book sales rankings.}
\seca{fitting}
\seca{Amazon}
\seca{4}

We next turn to the ranking in the Amazon.co.jp online book sales.
In this century of expanding online retail business,
the economic impact of internet retails has attracted much attention,
and there are studies using the sales rankings which appear
on the webs of online booksellers such as Amazon.com \cite{BSH,CG}.
We will study Amazon.co.jp, a Japanese counterpart of Amazon.com,
which seems to be not studied (and easier to access for the authors).
Basic structures of web pages for individual books are similar
for Amazon.com and Amazon.co.jp;
on a web page for a book there are the title, price and related information
such as shipping, brief description of the book, 
the sales ranking of the book,
customer reviews and recommendations.

We should note that
Amazon.co.jp, as well as Amazon.com, does not disclose
exactly how it calculates rankings of books.
In fact, there are observations \cite{Rosenthal}
that Amazon.com defines the rankings for the top sales 
in a rather involved way.
Therefore, it would be non-trivial and interesting if we could observe
in the data behaviors similar to those of our simple model such as
\eqnu{xclassicstationaryexpPareto}.
See \cite{HH073} for economic implications of the ranking numbers
in Amazon.co.jp.

According to observation,
Amazon.co.jp, as well as Amazon.com, updates their rankings once per hour, 
in contrast to the 2ch.net where the update procedure is instantaneous.
This implies a limit of short time observational precision of $1$ hour.
On the other hand, for the long time observations, 
we have to consider a fact that
the total number of books $N$ is not constant.
It is said that each year about $5\times 10^4$ books are published in Japan,
or about $5.7$ books per hour.
Certainly not all of the books are registered on Amazon.co.jp,
so the increase of $N$ per hour must be less than this value.
Speed of ranking change decreases in the very tail side of the listing,
and these practical changes in $N$ will 
affect validity of applying
\eqnu{xclassicstationaryexpPareto} to the data in the very tail regime 
of the ranking.
This gives a practical limit to long time analysis.
Fortunately, at ranking as far down as $6.5\times 10^5$,
we still observe about $200$ ranking change per hour, 
which makes an increase by $5.7$ books negligible,
so we expect a chance of applicability of our theory for long time data.

We will now summarize our results.
The plotted $77$ points in \figu{low2} show the result of our observation of
a Japanese book rankings data,
taken between the end of May, 2007 and mid August, 2007.
As seen in the figure, the ranking number falls very rapidly
near the top position (about $200$ thousands in $5$ days).
The solid curve is a least square
fit of these points to \eqnu{xclassicstationaryexpPareto}.
\begin{figure}[hbt]
\begin{center}
\includegraphics[scale=1.0]{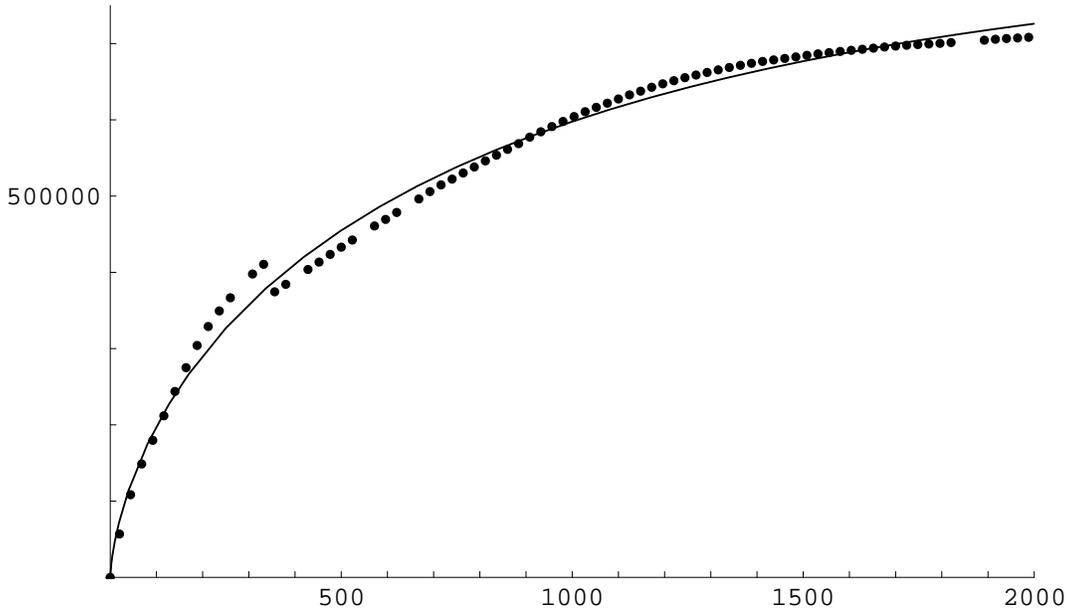}
\caption{A long time sequence of data from Amazon.co.jp.
The solid curve is a theoretical fit.
Horizontal and vertical axes are the hours and ranking, respectively.}
\figa{low2}
\end{center}
\end{figure}
Amazon.co.jp announces the total number of Japanese books in their list,
which is a few times $10^6$, but we suspect that this number includes
a large number of books which are registered but never sell
(so that we should discard in applying our theory).
Therefore in addition to $a$ and $b$ in \eqnu{xclassicstationaryexpPareto}
we include $N$ as a parameter to be fit from the data.
Also, Amazon does not disclose the exact point of sales of each book,
unlike 2ch.net where the exact jump time of a thread is recorded, 
so that the jump time to rank $1$ of a book is also a parameter.
The best fit for the parameter set $(N,a,b)$ is:
\[(N^*, a^*, b^*)=(8.57\times 10^5,3.939\times 10^{-4},0.6312),
\ \ \ (\sqrt{\chi^2/n_d}\simeq 1.4\times 10^4). \]

Incidentally, we note in \figu{low2} a small jump at about $300$ hours.
We suspect this as a result of inventory control
such as unregistering books out of print.
Obviously, these controls need man-power, so that they appear 
only occasionally,
making it a kind of unknown time dependent external source for our analysis.

All in all, we think it an impressive discovery that a simple 
formula as \eqnu{xclassicstationaryexpPareto} 
could explain the data for more than $2$ months.
Our way of extracting basic sociological
parameters such as the Pareto exponent $b$ from the ranking data
on the web has advantages over previous methods 
such as in \cite{CG}, in that because
the time development of the ranking of a book
is a result of sales of very large, $O(N)$, number of books,
the book moves on the ranking queue in a deterministic way
though each sale is a stochastic process.
The fluctuations of sales (randomness about
who buys what and when) are suppressed through a law-of-large-numbers
type mechanism \cite{HH071}.
By looking at the time development of the ranking of a single book,
we are in fact looking at the total sales of the books on the tail side 
of the book.

The theory of `long-tail economy' says \cite{longtail} that 
each product might sell only a little,
but because of the overwhelming abundance in the species of the products
the total sales will be of economic significance:
It is not any specific single book
but the total of books on the long-tail that matters.
Our analysis on
accumulated effect of products each with random and small sales,
is particularly suitable in analyzing the new and rapidly expanding 
economic possibility of online retails,
and moreover, is natural from the long-tail philosophy point of view.

Among the parameters  to be fit in the Pareto distribution
there is an exponent $b$
which is of importance in the studies of economy.
For example, in the case of distribution of incomes,
which is usually quantitatively analyzed by the Pareto distribution,
small $b$ means that a few people of high incomes hold most
of the wealth
(the so called `$20$--$80$ law' is a nickname for the Pareto distribution
with $b=1$),
while for large $b$ the society is more equal.
In the case of ranking of online booksellers, 
large $b$ means that there are many books
(books in the `long-tail' \cite{longtail} regime),
each of which do not sell much but
the total sales of which is significant,
further implying strong impacts of online retails to economy
\cite{CG,BSH},
while small $b$ 
favors dominance of traditional business model of `greatest hits'.
Our studies on the 2ch.net bulletin board and
 the Amazon.co.jp online bookseller
both consistently give the Pareto exponent  $b\simeq 0.6$.
Existing studies on online booksellers \cite{CG,BSH} adopt the value of 
$b=1.2$ and $b=1.148$, respectively.
These references also quote values from other studies, 
most of which satisfy $b>1$.
Note that \cite{Rosenthal} discovers, 
apparently based on extensive observations,
that in the long-tail regime
the sales are worse than in the head and intermediate regime,
and gives the Pareto exponent $b=0.4$ in the long-tail regime.
Our method gives the total effect of intermediate and long-tails,
so our value $b=0.6$ could be more or less consistent with the
observation of \cite{Rosenthal}.

\appendix
\section{Proof of equivalence of \protect\eqnu{e4} and \eqnu{bc2}.}
\seca{appendix}

Here we prove that, if $\sup_i f_i<\infty$,
 \eqnu{e4} and \eqnu{bc2} are equivalent,
under the equations \eqnu{e1} \eqnu{e2} \eqnu{e3} \eqnu{bc1} \eqnu{rho},
with positive constants $f_i$, $\rho_i$, $i=1,2,\cdots$.
(The extra condition on boundedness of $f_i$ is of course irrelevant
for the finite component cases.)

First assume \eqnu{e4}. Then with 
\eqnu{velocity}
we have
$\dsp v(0,t) =\sum_i f_i\, \rho_i$.
On the other hand, integrating \eqnu{e1} by $y$ from $0$ to $1$ 
and using \eqnu{e4} and \eqnu{bc1} we have
$v(0,t)\, u_i(0,t)=f_i\,\rho_i$.
The two equations imply \eqnu{bc2}.

Next assume \eqnu{bc2} and let 
\eqnb
\eqna{Ueqintu}
\vec{U}=(U_1,U_2,\cdots);\ \ 
 U_i(t)=\int_0^1 u_i(z,t)\,dz,\ i=1,2,\cdots.
\eqne
Then \eqnu{velocity} implies
\eqnb
\eqna{v0}
 v(0,t) =\sum_j f_j\,U_j(t).
\eqne
Integrating \eqnu{e1} by $y$ from $0$ to $1$, and using
\eqnu{bc1} \eqnu{bc2} \eqnu{v0},
we have
\eqnb
\eqna{evoleq}
 \diff{\vec{U}}{t}(t) = (A\,\vec{U})(t),\ \ t\ge 0,
\eqne
with
\eqnb
\eqna{generatorA}
 (A\,\vec{U})_i(t)=
\frac{\dsp f_{i}\rho_{i}}{\dsp \sum_{j}f_{j}\rho_{j}}\, \sum_j f_jU_j(t)
-f_{i} U_i(t), \  \ i=1,2,3,\cdots,\ t\ge0.
\eqne
The definition \eqnu{rho} implies $U_i(0)=\rho_i$,
hence $U_i(t)=\rho_i$, $t\ge 0$, is a solution, implying \eqnu{e4}.
Uniqueness of the solution to \eqnu{evoleq},
a linear differential equation with constant bounded coefficients,
proves \eqnu{e4}.

\end{document}